\documentclass[
    ,final            
  ]
  {aipproc}

\layoutstyle{6x9}

\newcommand{\numu}{\nu_{\mu}}
\newcommand{\numubar}{\bar{\nu}_{\mu}}

\newcommand{\Elo}{E_\mathrm{lo}}
\newcommand{\Ehi}{E_\mathrm{hi}}

\begin{document}
\title{Charged-Current Interaction Measurements in MiniBooNE}

\classification{11.80.Cr,13.15.+g,14.60.Lm,14.60.Pq}
\keywords{axial mass, charged current quasi-elastic, neutrino, 
MiniBooNE, Pauli blocking}

\author{Teppei Katori for the MiniBooNE collaboration}{
  address={Indiana University, Bloomington, IN}
}


\begin{abstract}
Neutrino oscillation is the only known phenomenon for physics beyond the
standard model. To investigate this phenomenon, the understanding of low energy 
neutrino scattering ($200<E_\nu<2000$ MeV) is the crucial task for 
high energy physicists. In this energy region, the charged current 
quasi-elastic (CCQE) neutrino interaction is the dominant process, and 
experiments require a precise model to predict signal samples. 
Using a high-statistics sample of muon neutrino CCQE events, 
MiniBooNE finds that a simple Fermi gas model, 
with appropriate adjustments, accurately characterizes the CCQE 
events on carbon. The extracted parameters include 
an effective axial mass, $M_A=1.23\pm 0.20$ GeV,
and a Pauli-blocking parameter, $\kappa = 1.019 \pm 0.011$.
\end{abstract}

\maketitle

\subsection{CCQE event selection in MiniBooNE}
\begin{figure}
\includegraphics[height=1.5in]{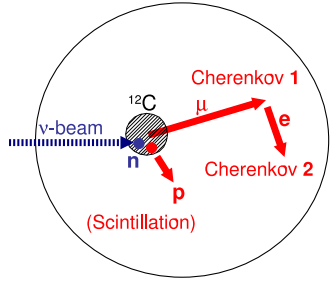}
\caption{Schematic figure of a CCQE interaction. 
The primary \v{C}erenkov light from the muon (\v{C}erenkov 1, first subevent) 
and subsequent \v{C}erenkov light from the decayed electron 
(\v{C}erenkov 2, second subevent) are used to tag the CCQE event. 
For most events, protons only emit scintillation light, 
and our selection is insensitive to this information.}
\label{fig:cartoon}
\end{figure}

The MiniBooNE detector, a spherical tank filled with mineral oil, detects 
\v{C}erenkov light from charged particles
\footnote{The detailed information of the Fermilab Booster neutrino beamline and 
the MiniBooNE neutrino detector are available elsewhere~\cite{MiniBooNE,CCQE}.}.
The identification of $\numu$CCQE interactions relies solely on 
the detection of the primary muon \v{C}erenkov light and 
the associated decay electron \v{C}erenkov light in 
these events (Fig.~\ref{fig:cartoon}):

\[
  \nu_\mu+n \rightarrow \mu^{-}+p \hspace{0.2in}
  \mu^{-} \rightarrow e^{-} + \nu_\mu + \bar{\nu_e}~~.
\]
By avoiding requirements on the outgoing proton kinematics, 
the selection is less dependent on nuclear models.
The scintillation light from the proton, 
although not used directly in the $\numu$CCQE analysis, 
is intensively studied in neutral current elastic 
scattering events at MiniBooNE~\cite{Cox}. 
A total of 193,709 events pass the MiniBooNE $\numu$CCQE selection 
criteria~\cite{CCQE} from 
$5.58\times10^{20}$ protons on target collected between 
August 2002 and December 2005. The cuts are estimated to be $35\%$ efficient 
at selecting $\numu$ CCQE events in a 500 cm radius, with a CCQE purity 
of $74\%$. 
The predicted backgrounds are: $74.8\%$ CC $1\pi^+$, $15.0\%$ 
CC $1\pi^0$, $4.0\%$ NC $1\pi^{\pm}$, $2.6\%$ CC multi-$\pi$, $0.9\%$ NC elastic, 
$0.8\%$ $\numubar$ CC $1\pi^-$, $0.8\%$ NC $1\pi^0$, $0.6\%$ $\eta$/$\rho$/$K$
production, and $0.5\%$ deep inelastic scattering and other events. 
Because pions can be absorbed via final state interactions in the target nucleus, 
a large fraction of the background events look like CCQE events in the MiniBooNE detector. 
``CCQE-like'' events, all events with a muon and no pions in the final state, 
are predicted to be $84\%$ of the sample after cuts.

\subsection{MiniBooNE CCQE events}
\begin{figure}
\includegraphics[height=2.4in]{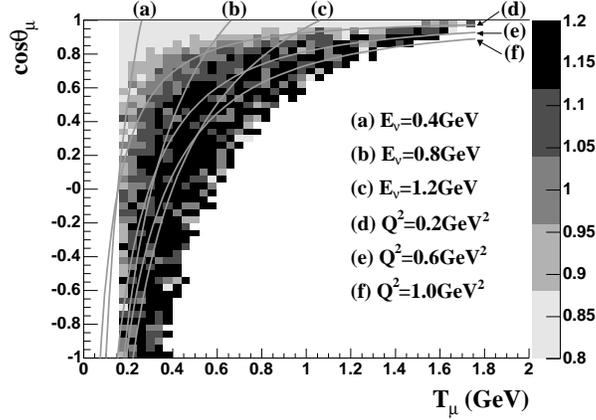}
\caption{Ratio of MiniBooNE $\numu$ CCQE data/simulation as a function of 
reconstructed muon angle and kinetic energy. The prediction 
is prior to any CCQE model adjustments; the $\chi^2/\mathrm{dof}=79.5/53$.
The ratio forms a 2D surface whose values are represented by the gray scale, 
shown on the right. If the simulation modeled the data perfectly, the ratio 
would be unity everywhere. Contours
of constant $E_\nu$ and $Q^2$ are overlaid.}
\label{fig:prl_fig1}
\end{figure}

Fig~\ref{fig:prl_fig1} shows the data-Monte Carlo (MC) 
ratio of CCQE events as a function of muon kinetic energy $T_{\mu}(GeV)$ 
and muon scattering angle $cos\theta_{\mu}$. 
Note the muon energy and muon scattering angle are the observables 
and the basis of all reconstructed kinematic variables in the $\nu_{\mu}$CCQE channel.
One can immediately see that the data-MC agreement is poor. 
There are 6 auxiliary lines:
 (a), (b), and (c) are equal neutrino energy lines, $0.4$, $0.8$, 
and $1.2GeV$ each, and (d), (e), and (f) are equal $Q^2$ lines,  
$0.2$, $0.6$, and $1.0 GeV^2$ each. The data-MC disagreement follows 
auxiliary lines of equal $Q^2$, not equal neutrino energy, 
this indicates that the data-MC disagreement is 
not due to the neutrino flux prediction, 
but due to the neutrino interaction prediction, 
because the former is a function 
of neutrino energy and the latter is a function of $Q^2$. 
So we assume that the data-MC disagreement comes from 
our neutrino interaction model and we adjust to the data. 
This is a critical task for MiniBooNE since the goal is to measure 
$\nu_e$CCQE events, but MC and all reconstruction tools must be 
reliable and tested in copious $\nu_{\mu}$CCQE events 
due to the blind analysis constraint on the $\nu_e$CCQE channel.

The data-MC disagreement is classified 
in 2 regions in this plane (Fig.~\ref{fig:prl_fig1}), 
\begin{enumerate}
\item data deficit at low $Q^2$ region, light gray band near the top left corner 
\item data excess at high $Q^2$ region, black band from the top right to the bottom left
\end{enumerate}
Since we are employing the Relativistic Fermi Gas (RFG) model~\cite{Smith-Moniz} 
in our MC, 
we wish to fix these problems within the RFG model.
The low $Q^2$ physics is usually controlled by nuclear model, 
so we want to tune the nuclear model, especially the strength of Pauli blocking, 
to fix region (1). This is justified because electron 
scattering data has not provided precise information for Pauli blocking 
in the low $Q^2$ region in terms of the RFG model~\cite{Butkevich}. 
For (2), we need to increase axial mass $M_A$ 
to increase the cross section at high $Q^2$. 
Here, the axial mass is understood as an experimental parameter 
in the axial form factor,

\begin{equation}
   F_A(Q^2) = \frac{g_A}{\left(1+\frac{Q^2}{M_A^2}\right)^2}~~,
\end{equation}

where $g_A$ is axial coupling constant (=$1.267$).
This treatment is also justified because elastic electron scattering cannot 
measure the axial mass precisely. Interestingly, 
the high axial mass is also observed by the K2K experiment in Japan~\cite{K2K}.

\subsection{Pauli blocking parameter $\kappa$}

\begin{figure}
\includegraphics[height=3.0in]{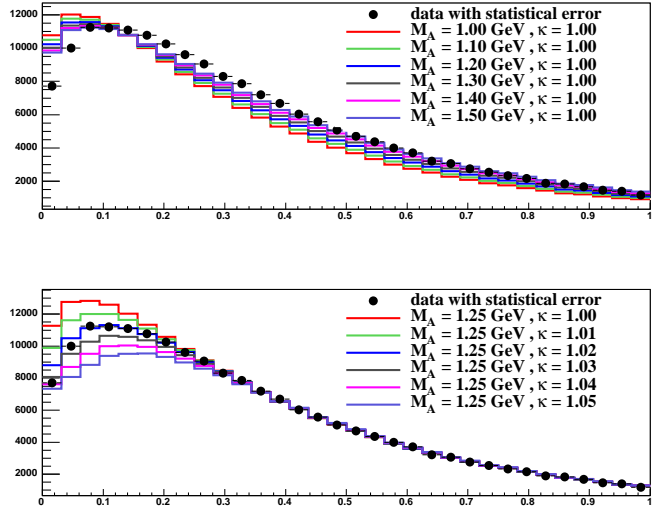}
\caption{Effect of $M_A$ and $\kappa$ variations 
on the MiniBooNE, flux integrated $Q^2$ distribution. 
The top plot shows various $M_A$ 
with fixed $\kappa$, and bottom plot shows various $\kappa$ 
with fixed $M_A$. Note, the $M_A$ variation has large impact at 
high $Q^2$ while the $\kappa$ variation has a significant 
impact only for $Q^2$ below $\sim 0.2GeV^2$.}
\label{fig:kappa}
\end{figure}

Currently, MiniBooNE is using the NUANCE neutrino 
interaction generator~\cite{NUANCE}. 
In NUANCE, CCQE interactions on carbon are modeled by the Relativistic Fermi Gas 
(RFG) model~\cite{Smith-Moniz}. To achieve our goal within the RFG model, 
we introduced a new parameter ``kappa'', $\kappa$,

\vspace{-0.1in}
\begin{eqnarray}
       \Ehi=\sqrt{p_F^2+M_n^2} \hspace{0.2in} 
       \Elo=\kappa(\sqrt{p_F^2+M_p^2} - \omega+E_B)
\end{eqnarray} 

where $M_n$ is the target neutron mass, $M_p$ is the outgoing proton mass, 
$P_F$ is Fermi momentum (=$220MeV$), $E_B$ is binding energy (=$34MeV$), 
and  $\omega$ is the energy transfer.
In the RFG model, $\Ehi$ is the energy of an initial nucleon on the Fermi 
surface and $\Elo$ is the lowest energy of an initial nucleon that leads to a 
final nucleon just above the Fermi momentum.
The function of parameter $\kappa$ is to squeeze down the 
phase space of the nucleon Fermi sea, especially when the energy transfer is small. 
From Fig.~\ref{fig:kappa}, one can see that 
this parameter controls the $Q^2$ distribution only in the low $Q^2$ region. 
This is quite complementary to the role of $M_A$, 
since $M_A$ mainly controls the $Q^2$ distribution in the high $Q^2$ region. 

We use these 2 parameters to perform a grid search to find the $\chi^2$ minimum. 
Here, we take into account all possible correlations between systematics 
by using the inverse of the full error matrix, not by adding systematics 
as pull terms.

\subsection{Fit result}

Finally, the parameters extracted from the MiniBooNE $\numu$ CCQE data are: 

\begin{eqnarray}
  M_A &=& 1.23 \pm 0.20 \hspace{0.1in} \mathrm{GeV}~; \\
  \kappa &=& 1.019 \pm 0.011~~.
\end{eqnarray}

Tab.~\ref{table:prl_tab1} shows the contributions to the systematic 
uncertainties on $M_A$ and $\kappa$. The detector model uncertainties
dominate the error in $M_A$ due to their impact on the energy and
angular reconstruction of CCQE events in the MiniBooNE detector. The
dominant error on $\kappa$ is the uncertainty in the $Q^2$ shape
of background events.

\begin{figure}
\includegraphics[height=2.6in]{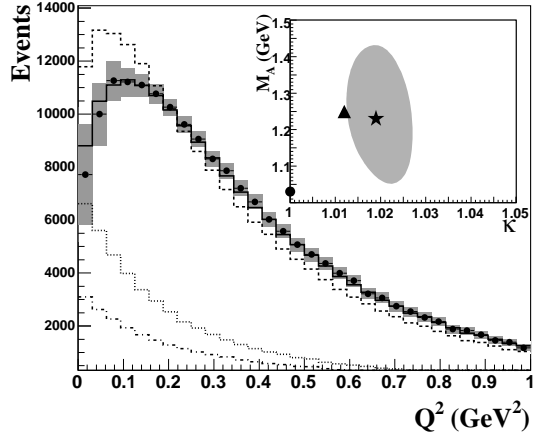}
\caption{Reconstructed $Q^2$ for $\numu$CCQE events including
systematic errors. The simulation, before (dashed) and after (solid) the fit, 
is normalized to data. 
The dotted (dot-dash) curve shows backgrounds that are not CCQE 
(not ``CCQE-like''). The inset shows the 1$\sigma$ CL contour for the best-fit 
parameters (star), along with the starting values (circle), and  
fit results after varying the background shape (triangle).}
\label{fig:prl_fig2}
\end{figure}

The result of this fitting, including all sources of systematic uncertainty,
is shown in Fig.~\ref{fig:prl_fig2}. 
Since the background error dominates at low $Q^2$, 
and it drives the large error bars at low $Q^2$. 
Note that, the shape uncertainty of the background, 
namely the $Q^2$ distribution shape uncertainty of $CC1\pi^{+}$ events, 
is not included in these error bands. 
From the data, we know that the predicted $Q^2$ shape 
of $CC1\pi^{+}$ events have large errors~\cite{Bonnie}. 
The extracted shape information from the data is implemented in our MC, 
and the fit is performed again. 
The result of those 2 fits, 
one using the MC predicted $CC1\pi^{+}$ distribution, 
and the other is using MC tuned on $CC1\pi^{+}$ data, 
are shown with the star and the triangle in the inserted plot 
in Fig.~\ref{fig:prl_fig2}. 
The difference is interpreted as a background shape uncertainty error 
and added to the extracted parameters. 
 
\begin{table}
\caption{Uncertainties in $M_A$ and $\kappa$ from the fit
to MiniBooNE $\numu$ CCQE data. The total error is not a simple quadrature
sum because of the correlation between the two parameters.}
\begin{tabular}{ccc}
\hline
error source&$\delta M_A$&$\delta \kappa$ \\
\hline
data statistics              &0.03             &0.003         \\ 
neutrino flux                &0.04             &0.003         \\
neutrino cross sections      &0.06             &0.004         \\
detector model               &0.10             &0.003         \\ 
CC $\pi^+$ background shape   &0.02             &0.007        \\ 
\hline
total error                  &0.20             &0.011        \\ 
\hline
\end{tabular}
\label{table:prl_tab1}
\end{table}

\begin{figure}
\includegraphics[height=2.5in]{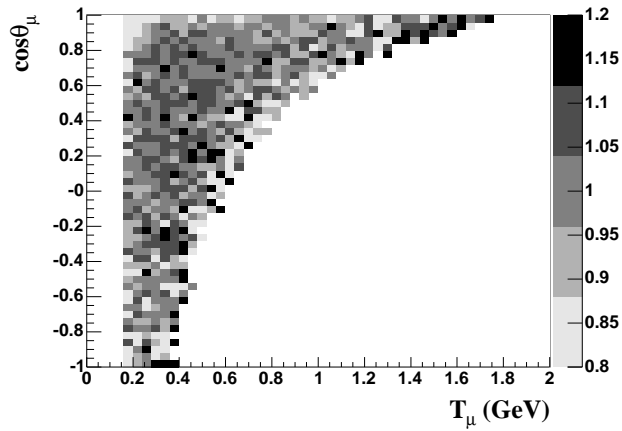}
\caption{Ratio of data/simulation as a function of muon kinetic energy and
angle after the CCQE model adjustments; the $\chi^2/\mathrm{dof}=45.1/53$.
Compare to Figure~\ref{fig:prl_fig1}.}
\label{fig:prl_fig3}
\end{figure}

\begin{figure}
\includegraphics[height=2.0in]{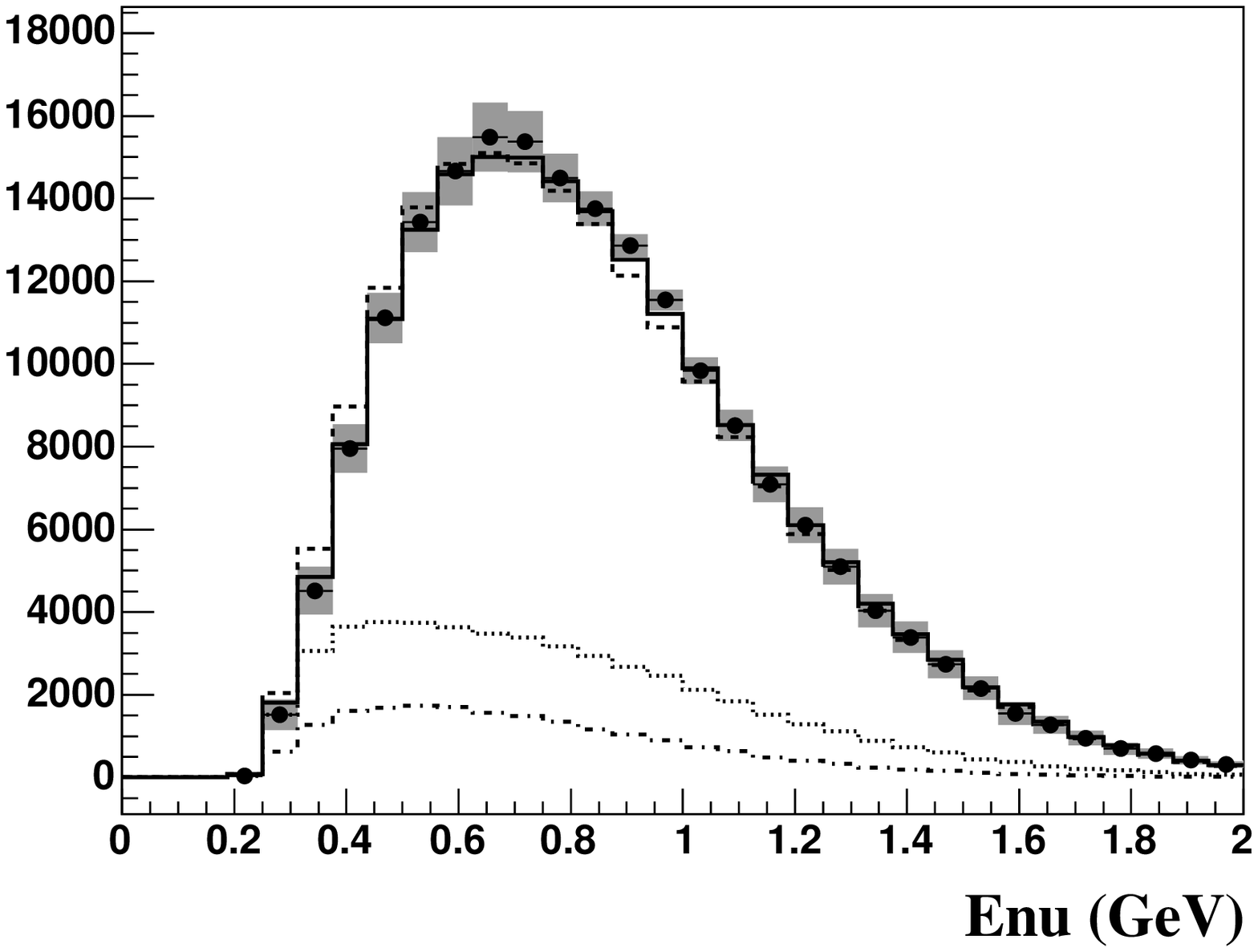}
\includegraphics[height=2.0in]{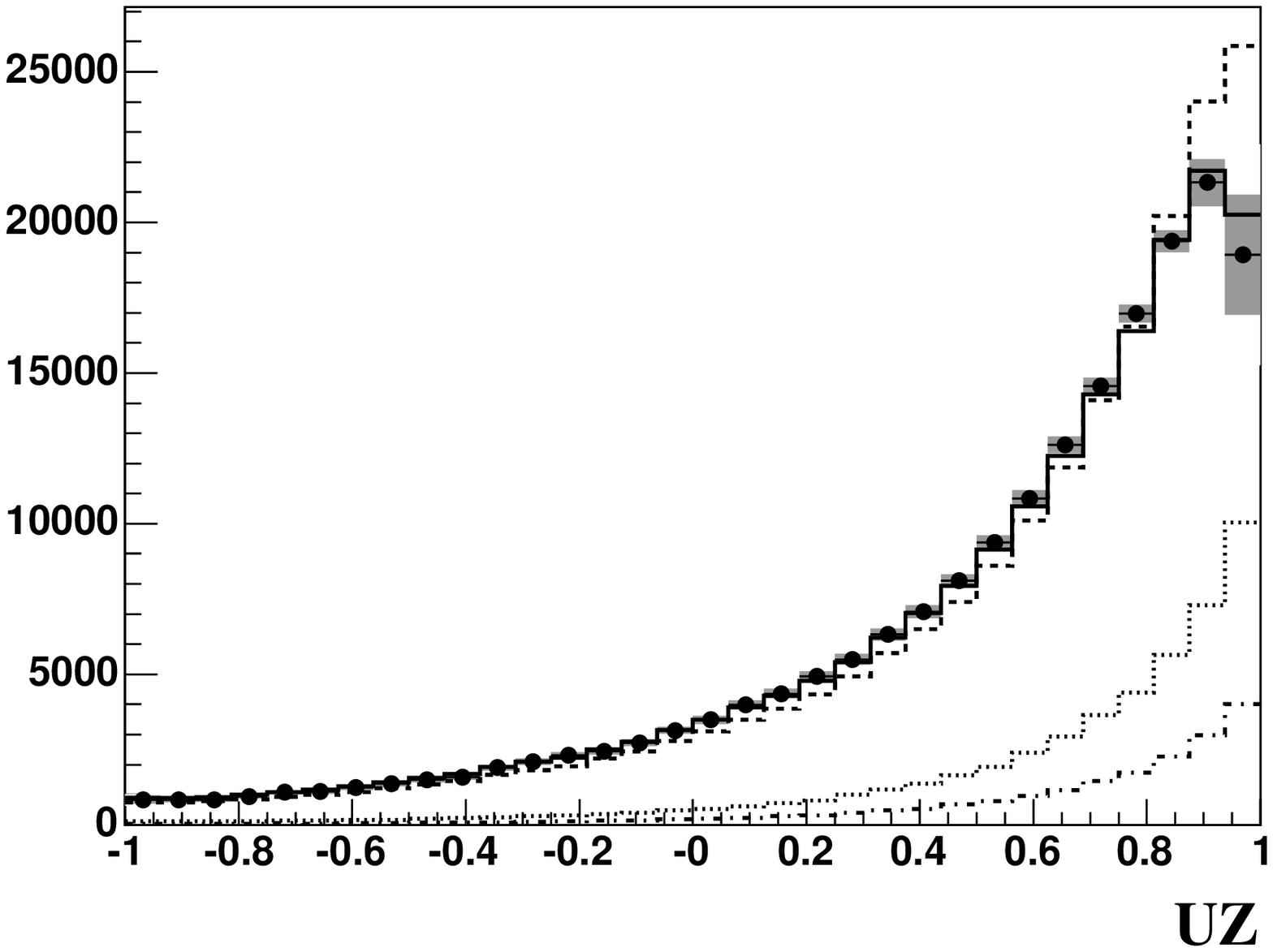}
\caption{Reconstructed muon neutrino energy and 
measured muon scattering angle, line notations are 
the same as Fig~\ref{fig:prl_fig2}.}
\label{fig:prl_fig2_other}
\end{figure}

Fig.~\ref{fig:prl_fig3} shows the agreement between data and 
simulation after incorporating the $M_A$ and $\kappa$ values from 
the $Q^2$ fit to MiniBooNE $\numu$CCQE data. 
Comparing to Fig.~\ref{fig:prl_fig1}, 
the improvement is substantial and the data are well-described throughout 
the kinematic phase space. Since the whole kinematic space is fixed, 
not surprisingly, 
all of the individual kinematic variables exhibit good data-MC agreement.  
Fig.~\ref{fig:prl_fig2_other} shows that data and MC agree well within error bars
for reconstructed muon neutrino energy and muon scattering angle.

\begin{figure}
\includegraphics[height=2.0in]{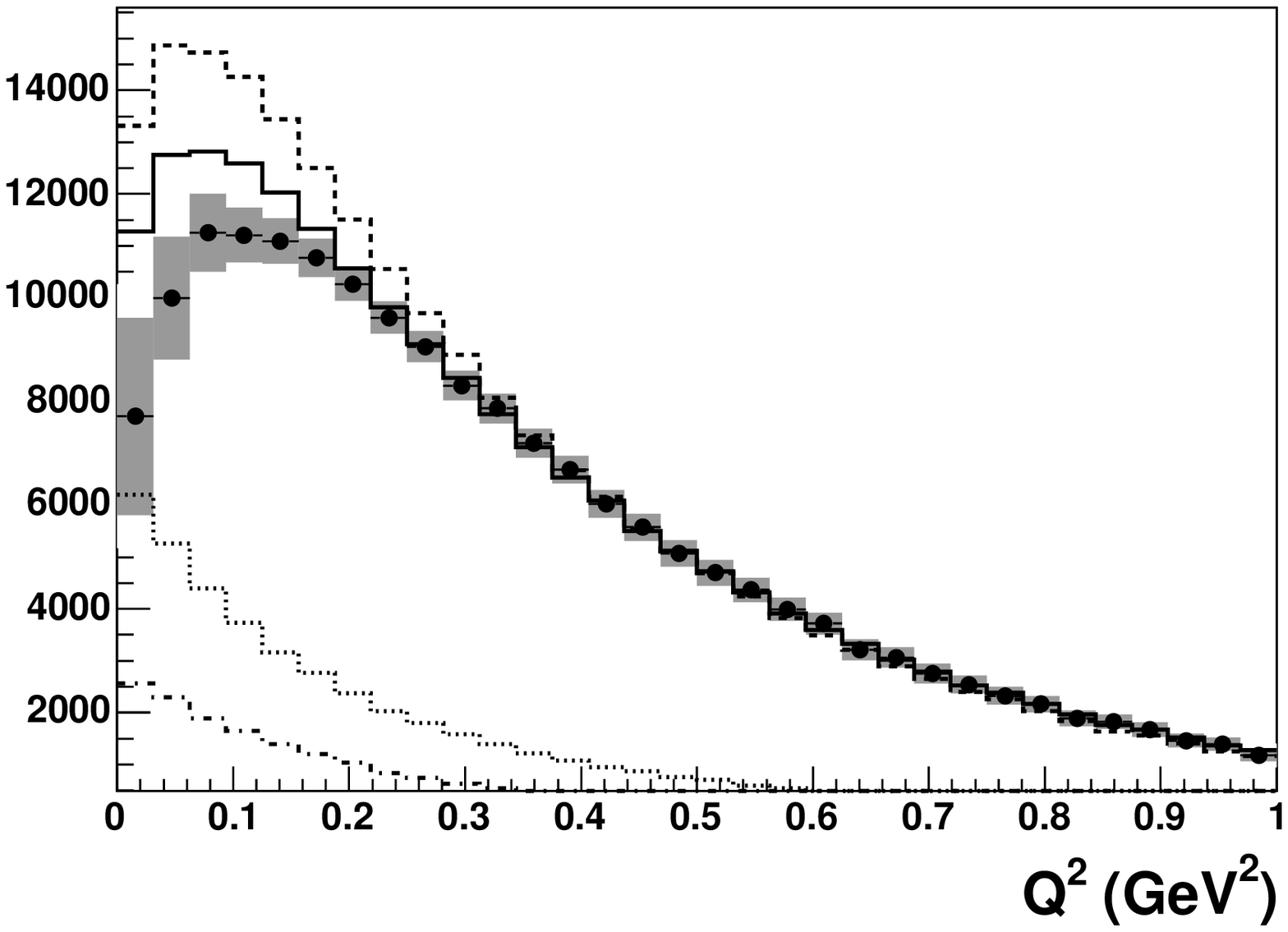}
\includegraphics[height=2.0in]{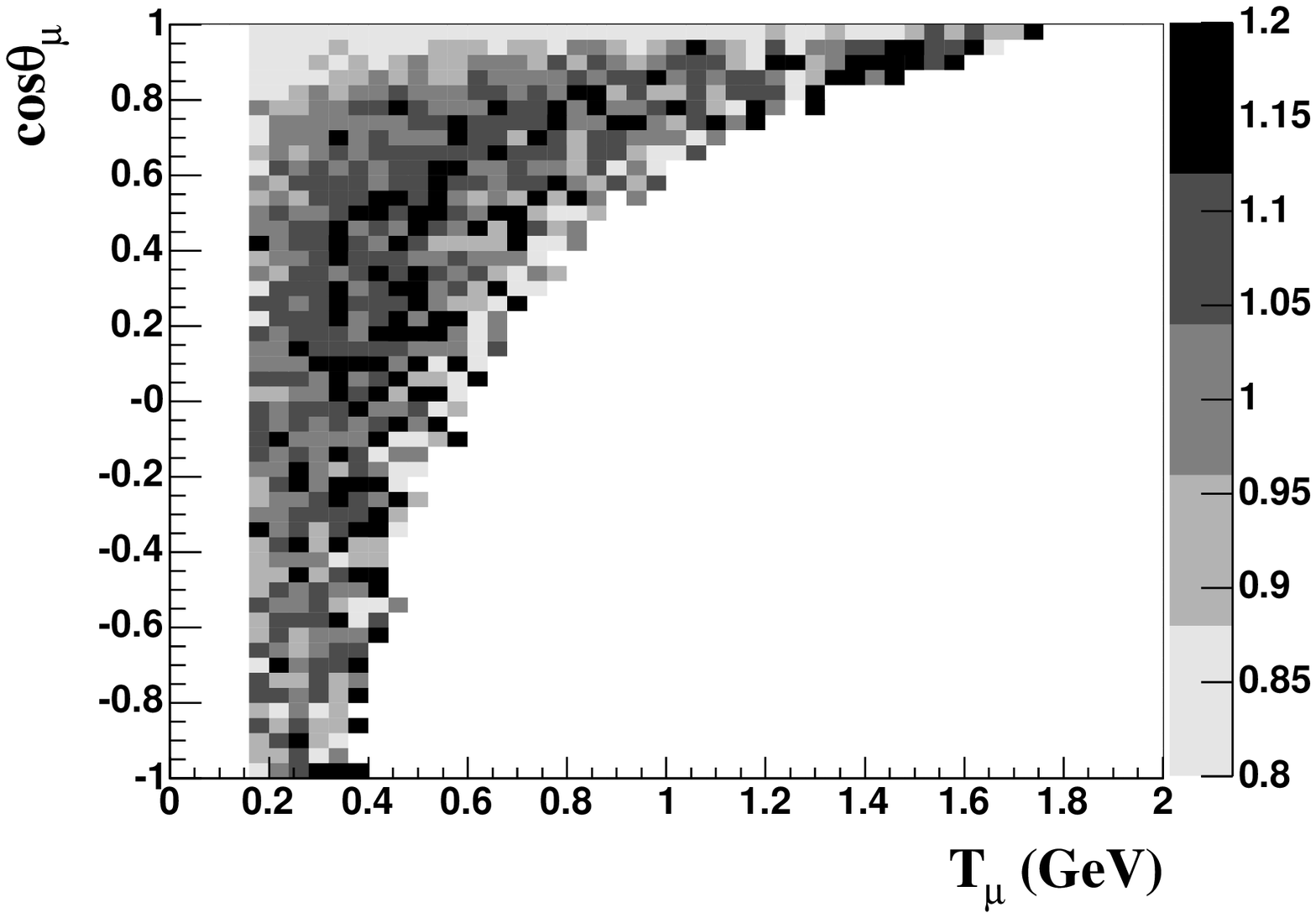}
\caption{Reconstructed $Q^2$ for $\numu$ CCQE events and data-MC ratio 
in the kinematic plane. 
The left plot is the analogy of Fig.~\ref{fig:prl_fig2}, 
and the right plot is the analogy of Fig.~\ref{fig:prl_fig3}
but the fit is performed using $M_A$ only, with fixed $\kappa$ 
(=1.0, no enhanced Pauli blocking).}
\label{fig:prl_fig2_maonly}
\end{figure}

\vspace{0.2in}
In general, varying $M_A$ allows us to reproduce the high $Q^2$ behavior of
the observed data events. A fit for $M_A$ above $Q^2>0.25$ GeV$^2$ yields 
consistent results, $M_A = 1.25 \pm 0.12$ GeV (Fig.~\ref{fig:prl_fig2_maonly}). 
However, fits varying only $M_A$ across the entire 
$Q^2$ range leave considerable disagreement at low $Q^2$. 
This data-MC disagreement at low $Q^2$ would eventually reflect in 
data-MC disagreement in reconstructed neutrino energy, 
because data-MC disagreement in $Q^2$ spreads out in the kinematic plane 
and would affect the energy reconstruction across a wide region. 
The Pauli-blocking parameter $\kappa$ 
is instrumental here, enabling this model to match the behavior of the data 
down to $Q^2=0$.

\subsection{Anti-neutrino CCQE preliminary result}
\begin{figure}
\includegraphics[height=2.2in]{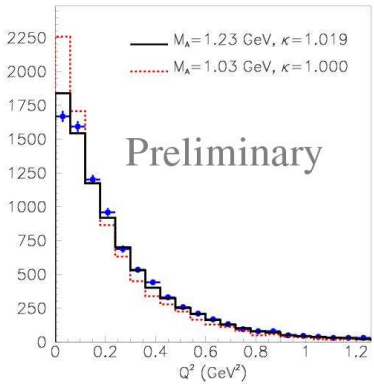}
\caption{Reconstructed $Q^2$ for $\numubar$CCQE events, including
statistics error only. The simulation, old model (dashed) and new model (solid) 
is normalized to data.}
\label{fig:anti}
\end{figure}

Finally, we tested the modified RFG model 
in a new sample of MiniBooNE antineutrino data 
(for other preliminary results for antineutrino run, see~\cite{Van}). 
If our assumption is correct, 
this RFG model should also succeed in modeling $\numubar$CCQE events. 
The result is shown in Fig.~\ref{fig:anti}. 
Although statistics are low, one can tell 
the new model will describe the features of this data better than the original model.  

\vspace{0.2in}
In summary, taking advantage of the high-statistics MiniBooNE $\numu$ CCQE  data, 
we have extracted values of an effective axial mass parameter, 
$M_A=1.23 \pm 0.20$~GeV, and a Pauli-blocking parameter, 
$\kappa=1.019\pm 0.011$, achieving substantially improved agreement with the 
observed kinematic distributions in this data set. 

The $M_A$ value reported here should
be considered an ``effective parameter'' in the sense that it may be 
incorporating nuclear effects not otherwise included in the RFG model. Future
efforts will explore how the value of $M_A$ extracted from the MiniBooNE 
data is altered  upon replacement of the RFG model with more advanced nuclear 
models~\cite{new-model}.



\end{document}